\documentclass[twocolumn,showpacs,prl,superscriptaddress]{revtex4}
\usepackage{graphicx}
\usepackage{dcolumn}
\usepackage{amsmath}
\usepackage{amssymb}

\makeatletter
\def\btt#1{\texttt{\@backslashchar#1}}
\DeclareRobustCommand\bblash{\btt{\@backslashchar}} \makeatother

\renewcommand{\Re}{\mathop\mathrm{Re}\nolimits}
\renewcommand{\Im}{\mathop\mathrm{Im}\nolimits}

\begin{document}

\title{Odd-frequency Cooper Pairs and Their Effects on Surface Impedance}

\author{Yasuhiro Asano}
\affiliation{Department of Applied Physics and Center for Topological Science \& Technology,
Hokkaido University, Sapporo 060-8628, Japan}

\author{Alexander A. Golubov}
\affiliation{Faculty of Science and Technology and MESA+ Institute of Nanotechnology,
University of Twente, 7500 AE, Enschede, The Netherlands}

\author{Yakov V. Fominov}
\affiliation{L.~D.~Landau Institute for Theoretical Physics, RAS, 119334 Moscow, Russia}

\author{Yukio Tanaka}
\affiliation{Department of Applied Physics, Nagoya University, Nagoya 464-8603, Japan}

\date{\today}

\begin{abstract}
We discuss dynamic response of odd-frequency Cooper pairs to electromagnetic field.
By using the quasiclassical Green function method, we calculate the 
impedance ($Z=R-iX$) of a normal metal thin film which covers a superconductor.
In contrast to the standard relation (i.e., $R\ll X$),
the impedance in spin-triplet proximity structures shows
anomalous behavior (i.e., $R>X$) in the low frequency limit.
This unusual relation is a result of the penetration of odd-frequency pairs into the
normal metal and reflects the \textsl{negative Cooper pair density}.
\end{abstract}


\maketitle

Exotic superconductors with a gap function $\Delta(\boldsymbol{k},\omega_n)$
odd in Matsubara frequency $\omega_n$ are called odd-frequency superconductors.
Mysterious behavior of odd-frequency Cooper pairs is an intriguing issue
in the physics of superconductivity and superfluidity \cite{berezinskii,balatsky92}.
An odd-frequency superconductor might not exhibit essential properties of superconductivity
such as the gapped quasiparticle spectrum and the standard diamagnetic Meissner effect.
Unfortunately, the odd-frequency pairing state has never been yet detected experimentally
in bulk materials. Odd-frequency Cooper pairs themselves, however,
may exist in superconducting proximity structures. It was theoretically predicted in
ferromagnet/superconductor junctions that spin-mixing due to spin-dependent potential should
generate odd-frequency spin-triplet $s$-wave pairs \cite{bergeret,eschrig,braude,ya07sfs,linder}.
Manifestations of triplet pairs were recently observed experimentally as a long-range
Josephson coupling across ferromagnets \cite{Keizer,Anwar,Robinson,Khaire}.
An alternative way of creating odd-frequency pairs was suggested in proximity structures
involving a normal metal attached to an odd-parity spin-triplet superconductor
that belongs to the conventional even-frequency symmetry class.
The parity-mixing due to inhomogeneity produces the odd-frequency pairs
even in this case \cite{tanaka07e}.

Although the experiments~\cite{Keizer,Anwar,Robinson,Khaire} catch a sign of 
odd-frequency pairs and
theories \cite{yt04,yt05r,ya06,ya07,fominov} have predicted 
unusual properties of spin-triplet superconducting junctions,
we have never had clear scientific evidences of odd-frequency pairs.
This is because physical values focused in these studies include 
only {\em indirect} information of the frequency symmetry.
The present paper shows that the surface impedance
is a particular physical value which {\em directly} reflects the frequency symmetry of a Cooper pair.
We predict anomalous features of a surface impedance which should 
make possible to detect odd-frequency pairs experimentally.

\begin{figure}[tbh]
\begin{center}
\includegraphics[width=8cm]{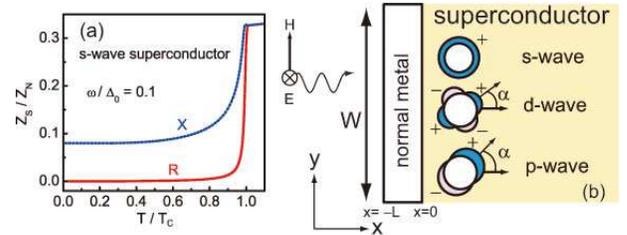}
\end{center}
\caption{(color online). (a)~Typical behavior of the surface impedance
in a conventional $s$-wave superconductor.
(b)~Proximity structure under consideration, where a
thin film of a normal metal covers a superconductor.
}
\label{fig1}
\end{figure}
Surface impedance $Z=R-iX$ reflects the dynamic response of Cooper pairs
to electromagnetic field \cite{mattice,nam}.
The surface resistance, $R$, corresponds to resistance due to normal electrons.
In conventional superconductors, $R$ drops exponentially with
decreasing temperature below superconducting transition temperature $T_c$.
The reactance, $X$, represents power loss of electromagnetic field due to Cooper pairs.
Typical temperature dependence of the surface impedance in a conventional superconductor
is shown in Fig.~\ref{fig1}(a).
The positive amplitude of the Cooper pair density guarantees a robust relation $R\ll X$
at low temperatures.
The validity of the relation $R<X$, however, is questionable for odd-frequency
Cooper pairs because the odd-frequency symmetry and \textit{negative pair density}
are inseparable from each other according to the standard theory of superconductivity \cite{agd}.
In the present paper, we conclude that the unusual relation $R>X$ can be observed 
in the presence of odd-frequency Cooper pairs at low frequency and at low temperature.

Let us consider a bilayer of a superconductor and a thin normal metal film
as shown in Fig.~\ref{fig1}(b), where $W$ is the size of the superconductor
and $L$ is the thickness of the normal metal.
We mainly consider spin-triplet $p$-wave symmetry in the superconductor and compare
results with those in spin-singlet $s$- and $d$-wave symmetries.
In all cases, the superconductor belongs to the conventional even-frequency symmetry class.
The even- and odd-frequency pairs penetrate into the normal metal for 
the spin-singlet and spin-triplet cases, respectively~\cite{tanaka07,yt04,yt05r}. 
To calculate the complex conductivity in the normal metal, we first solve the quasiclassical Usadel
equation~\cite{usadel} in the standard $\theta$-parametrization~\cite{tau1},
$\hbar D \nabla^2 \theta(\boldsymbol{r},\epsilon) +
2i \epsilon \sin \theta(\boldsymbol{r},\epsilon)=0$,
where $D$ is the diffusion constant of the normal metal and
$\epsilon$ is the quasiparticle energy measured from the Fermi level.
In what follows, we consider the normal-metal/superconductor (NS) junction
in two dimension as shown in Fig.~\ref{fig1}(b).
We choose $W\gg L$ so that we can neglect $y$ dependence of $\theta$.
In the $p$-wave case, we assume that a spin-triplet Cooper pair
consists of two electrons with opposite spin directions ---
this assumption does not break generality of the following argument.
The Usadel equation is supplemented by the boundary condition at the NS
interface which depends on the pairing symmetry of the superconductor \cite{nazarov,yt03-1,yt04},
\begin{align}
&\left.\frac{\partial \theta(x,\epsilon)}{\partial x}\right|_{x=0}
= \frac{1}{L}\frac{R_D}{R_B} \frac{\langle F \rangle}{T_B}, \label{naz}\\
\langle F \rangle =& \int_{-\pi/2}^{\pi/2} \!\!\! d\gamma
\frac{ T_N \cos\gamma  (f_s \cos\theta_0 - g_s \sin\theta_0)}
{(2- T_N) + T_N ( g_s \cos\theta_0 + f_s \sin\theta_0)}, \label{deff}
\end{align}
where $\gamma$ is the incident angle of
a quasiparticle measured from the $x$ axis,
$\theta_0=\theta(x=0,\epsilon)$, and $R_D$ is the resistance of the normal metal.
At the NS interface, a potential barrier described by $\mathrm v_0\delta(x)$ suppresses
the transmission probability $T_B = \int_0^{\pi/2}\!\!\! d\gamma \cos\gamma\; T_N$
with $T_N= \cos^2\gamma/(z_0^2 + \cos^2\gamma)$, $z_0=\mathrm v_0/(\hbar v_F)$,
and $v_F$ being the Fermi velocity. As a result,
the resistance of the NS interface becomes $R_B=[(e^2/\pi\hbar)(k_FW/\pi)T_B]^{-1}$.
The Green function in the superconductor depends on $\gamma$ and the orientation angle
$\alpha$ in Fig.~\ref{fig1}(b) as
$g_\pm = \epsilon / \sqrt{ \epsilon^2 - |\Delta_\pm|^2}$
and $f_\pm = i\Delta_\pm / \sqrt{\epsilon^2-|\Delta_\pm|^2}$, where
$\Delta_\pm=\Delta\Psi(\gamma_\pm)$ with $\Delta$ being the amplitude of the
pair potential,  $\gamma_+=\gamma-\alpha$ and $\gamma_-=\pi -\gamma -\alpha$.
The form factor $\Psi(\gamma)$ characterizes the pairing symmetry as
$\Psi(\gamma)=1$, $\cos \gamma$, and $\cos 2\gamma$ for the $s$-, $p$-,
and $d$-wave symmetries, respectively.
At a particular orientation angle, $p$-wave with $\alpha=0$, $p$-wave with $\alpha=\pi/2$, $d$-wave
with $\alpha=0$, and $d$-wave with $\alpha=\pi/4$ correspond to the $p_x$-, $p_y$-, $d_{x^2-y^2}$-, and
$d_{xy}$-symmetry, respectively.
In Eq.\ (\ref{deff}), $g_s=(g_++g_-)/\xi$ with
$\xi=1+g_+g_-+f_+f_-$, and $f_s=(f_++f_-)/\xi$ for the spin-singlet pairing symmetry,
while $f_s= i(f_+g_- - f_-g_+)/\xi$ for the spin-triplet one \cite{yt03-1,yt04}.
For the chiral $p$-wave symmetry, when $\Psi(\gamma)=e^{i\gamma}$, $\theta$ obeys a boundary condition which is slightly different
from Eq.\ (\ref{deff}) \cite{yt05r}.
At the outer surface of the normal metal, we require $ \partial_x \theta(x,\epsilon)|_{x=-L}=0$.
The normal and anomalous retarded Green functions
are obtained
as $g(x,\epsilon)=\cos\theta(x,\epsilon)$ and $f(x,\epsilon)=\sin\theta(x,\epsilon)$,
respectively.

Having found the Green functions, we can calculate the local complex conductivity 
$\sigma_{\textrm{N}}(x,\omega) = \sigma_1+i \sigma_2$ which
can be represented only by the retarded Green
function as~\cite{FGH}
\begin{align}
\frac{\sigma_1(x,\omega)}{\sigma_0} &= \frac{1}{2\hbar\omega}
 \int_{-\infty}^{\infty} \!\!\!\!\!\! d\epsilon \left[J(\epsilon+\hbar\omega)-J(\epsilon)
\right] K_1,\label{sig1y}\\
\frac{\sigma_2(x,\omega)}{\sigma_0} &= \frac{1}{2\hbar\omega}
 \int_{-\infty}^{\infty}\!\!\!\!\!\! d\epsilon \left[
J(\epsilon+\hbar\omega) K_2  + J(\epsilon) K_3 \right],\label{sig2y}\\
K_1 &= f_I(\epsilon)f_I(\epsilon+\hbar\omega) + g_R(\epsilon)g_R(\epsilon+\hbar\omega),\label{k1}\\
K_2 &= f_R(\epsilon)f_I(\epsilon+\hbar\omega) - g_I(\epsilon)g_R(\epsilon+\hbar\omega),\label{k2}\\
K_3 &= f_R(\epsilon+\hbar\omega)f_I(\epsilon) - g_I(\epsilon+\hbar\omega)g_R(\epsilon),\label{k3}
\end{align}
with $J(\epsilon)= \tanh\left({\epsilon}/{2k_B T}\right)$,
$g_R(\epsilon) = \Re \left[ g(x,\epsilon) \right]$, $g_I(\epsilon) =
\Im \left[ g(x,\epsilon) \right]$,
$f_R(\epsilon) = \Re \left[ f(x,\epsilon) \right]$, and
$f_I(\epsilon) = \Im \left[ f(x,\epsilon) \right]$.
The local impedance in the normal metal is calculated
from the complex conductivity as
\begin{equation}
Z_{\textrm{N}}(x,\omega)=R_{\textrm{N}}-iX_{\textrm{N}}= (1-i) \sqrt{\frac{\hbar\omega}{\Delta_0}
\frac{\sigma_0}{\sigma_{\textrm{N}}(x,\omega)}}{Z_\textrm{0}},\label{zns}
\end{equation}
where $Z_{\textrm{0}}\equiv \sqrt{2\pi \Delta_0/\sigma_0c^2\hbar}$,
$\Delta_0$ is the amplitude of pair potential at $T=0$, and $\sigma_0$
is the Drude conductivity in the normal metal. 
The surface impedance in Fig.~\ref{fig1}(a) is calculated from the above formula,
taking the Green function of a uniform $s$-wave superconductor.

The theory basically includes three junction parameters:
the thickness of normal metal $L$, strength of the potential barrier $z_0$, and
the diffusion constant in the normal metal $D$, which are converted to
$R_D/R_B$, $E_\mathrm{Th}=\hbar D/L^2$, and $z_0$ with $E_\mathrm{Th}$ being the
Thouless energy. 
As we discuss later on, the results shown below depend on
these parameters only quantitatively.
Thus we fix them at
$R_D/R_B=1$, $E_\mathrm{Th}=0.5\Delta_0$, and $z_0=2.5$ throughout this paper,
which leads to $T_B \approx 0.1$ and $L \approx 2.7 \xi_{T_c}$ with 
$\xi_{T_c}=\sqrt{\hbar D/2\pi k_B T_c}$ being the coherence length.
We add a small positive imaginary part $i\epsilon_{\lambda}$ to energy
to provide the retarded causality and to
obtain reasonable convergence of the integration in Eqs.\ (\ref{sig1y}) and (\ref{sig2y}).
In real junctions, $i\epsilon_{\lambda}$ stems from inelastic scatterings of a
quasiparticle due to phonon or electron-electron interaction.
We fix $\epsilon_{\lambda}$ at $0.01 \Delta_0$.

\begin{figure}[tbh]
\begin{center}
\includegraphics[width=8cm]{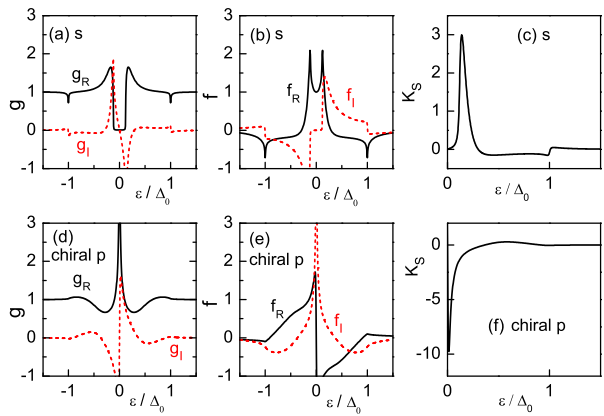}
\end{center}
\caption{(color online).
The retarded Green functions at the surface of the normal metal 
for the $s$-wave (a)-(b)
 and the chiral $p$-wave symmetry (d)-(e).
The solid and broken lines represents the real and imaginary part of the
Green function, respectively.
In (c) and (f), $K_s$ is the spectral pair density and is corresponding to the kernel of the
imaginary part of complex conductivity at $\omega\to 0^+$.
}
\label{fig2}
\end{figure}

First, we compare the retarded Green functions at the surface of the normal metal ($x=-L$)
for the $s$-wave symmetry in Fig.~\ref{fig2} (a)(b) with those for the chiral $p$-wave in (d)(e).
The real part of the normal Green function $g_R(\epsilon)$ is the quasiparticle density of
states normalized by the normal density of states at the Fermi level.
The normal Green function always satisfies $g(-\epsilon)=\left[ g(\epsilon) \right]^\ast$
irrespective of pairing symmetry as shown in (a) and (d).
In the $s$-wave case, $g_R(\epsilon)$ shows a minigap for
$|\epsilon| < {\Delta}_N\sim 0.15 \Delta_0$.
In chiral $p$-wave case, $g_R$ in (d) shows a large peak around $\epsilon=0$,
which can be understood as the penetration of midgap Andreev resonant
states into the normal metal. 
The formation of midgap Andreev resonant states due to a sign change of the pair 
potential is a common feature at surfaces of unconventional superconductors~\cite{tanaka95}.
The penetration of such states into a normal metal is possible only for
spin-triplet junctions~\cite{yt04,ya06}.
The feature of the anomalous function 
$f_R(\epsilon)$ depends on the
pairing symmetry of the superconductor.
The results in (b) show that $f_R(\epsilon)$ is an even function of $\epsilon$ and
$f_I(\epsilon)$ is an odd function of $\epsilon$ (i.e.,
$f(-\epsilon)=\left[ f(\epsilon) \right]^\ast$). This relation
means the presence of the even-frequency Cooper pairs in the normal metal
and always holds for spin-singlet NS structures.
The results in (e), however, show that
$f_R(\epsilon)$ is an odd function of $\epsilon$ and $f_I(\epsilon)$
is an even function of $\epsilon$
(i.e., $f(-\epsilon)=-\left[ f(\epsilon) \right]^\ast$) \cite{tanaka07}.
This relation represents
the penetration of odd-frequency Cooper pairs into the normal metal and
holds for spin-triplet NS structures.

\begin{figure}[tbh]
\begin{center}
\includegraphics[width=8cm]{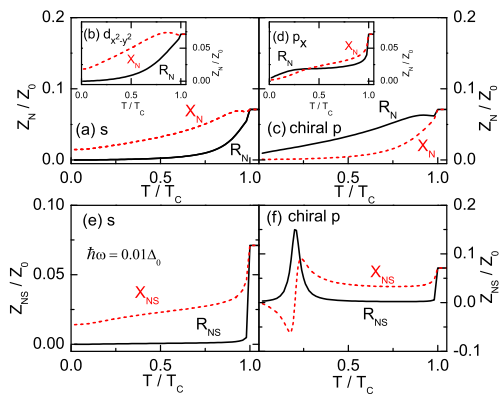}
\end{center}
\caption{(color online).
Local impedance at a surface of the normal metal (i.e., $x=-L$) versus temperature 
at $\hbar\omega = 0.01 \Delta_0$ for (a)~$s$-wave, (b)~$d_{x^2-y^2}$-wave, 
(c)~chiral $p$-wave, and (d)~$p_x$-wave symmetry.
The impedance of a NS bilayer versus temperature 
for (e) $s$-wave and (f) chiral $p$-wave symmetry.
}
\label{fig3}
\end{figure}

Secondly, we show the local impedance at the surface of the normal metal 
$Z_{\textrm{N}}(-L,\omega)$ as a function of 
temperature at a small
$\hbar\omega$ fixed at $0.01\Delta_0$
in Fig.~\ref{fig3}.
Here the local impedance $Z_{\textrm{N}}(-L,\omega)$ approximately describes 
the impedance of the normal metal because $\sigma_{\textrm{N}}(x,\omega)$ depends on 
$x$ weakly for $L \approx \xi_{T_c}$.  
In the $s$-wave symmetry, the results show conventional behavior totally consistent with
those in Fig.~\ref{fig1}(a). Namely, both $R_{\textrm{N}}$ and $X_{\textrm{N}}$ decrease
 with decreasing $T$
at low temperatures with keeping the relation $R_{\textrm{N}}\ll X_{\textrm{N}}$.
The results for $d_{x^2-y^2}$-symmetry in (b) show qualitatively the same feature
as those in the $s$-wave case.
For $d_{xy}$- and $p_y$-symmetries,
a relation $R= X$ always holds irrespective of $\omega$ and $T$ (results are not shown)
because there is no proximity effect in the normal metal \cite{yt03-1,ya01}.
In this case, the impedance reflects only the skin effect of the normal metal.
In the chiral $p$-wave symmetry in (c), on the other hand, $R_{\textrm{N}}$ become larger 
than $X_{\textrm{N}}$ 
for $T < T_c$. 
A similar behavior can be seen also in the $p_x$-wave symmetry
for $T<T^\ast \sim 0.4 T_c$ as shown in (d).
The impedance clearly exhibits different characteristic behaviors depending on 
the pairing symmetry of the superconductor.
In Figs. (c) and (d),
$\sigma_2$ changes its sign to negative.
It is easy to confirm that the sign change of $\sigma_2$ results in 
$R_N>X_N$ in Eq.\ (\ref{zns}). Thus $R_N>X_N$ is a robust and universal 
property of odd-frequency Cooper pair.

To understand characteristic feature of complex conductivity at small $\omega$,
we analyze the spectral pair density defined by
\begin{equation}
K_s(\epsilon) = f_R(\epsilon)f_I(\epsilon)-g_R(\epsilon)g_I(\epsilon)
=\textrm{Im}f^2(\epsilon),\label{ks}
\end{equation}
which appears in the integrand of $\sigma_2$ in Eq.\ (\ref{sig2y})
at very small $\omega$. 
The spectral pair density contains full information about
the frequency symmetry of Cooper pairs.
In the $s$-wave case, $K_s$ is mostly positive for $\epsilon>0$ and
has a positive peak around $\epsilon={\Delta}_N$ as shown in Fig.~\ref{fig2}(c).
Since $K_s$ is an odd function of $\epsilon$ according to its definition,
we only show the results for $\epsilon>0$.
On the other hand in the chiral $p$-wave symmetry as shown in Fig.~\ref{fig2}(f), 
$K_s$ is mostly negative
for $\epsilon>0$ and has a large negative peak near $\epsilon=0$ due to 
the odd-frequency symmetry.
The Cooper pair density in the normal metal 
$n_s =\int_{-\infty}^\infty d\epsilon J(\epsilon) K_s(\epsilon)$ 
is positive for the $s$-wave case and is negative for the chiral $p$-wave one.
As a result, $\sigma_2$ is positive for the $s$-wave case
and is negative for the chiral $p$-wave case. 
Therefore we conclude that the negative $\sigma_2$ is the \textit{direct} consequence of
the odd-frequency symmetry of Cooper pairs. 
This argument generally holds when $\hbar\omega$ 
is smaller than all energy scales relating to the proximity effect. The smallest 
energy scale at the present calculation is the peak width of the Green 
function shown in Fig.~\ref{fig2}(a)-(d).
Indeed we have confirmed that $T^\ast$ in Fig.~\ref{fig3}(d) increases to $T_c$ 
with decreasing $\omega$.  
The conclusion does not change even if we vary the junction parameters such as 
$R_D/R_B$, $E_\mathrm{Th}$, and $z_0$,
since these parameters do not affect the frequency symmetry of Cooper pairs.

Next, we briefly discuss physics behind the negative $\sigma_2$ based on a phenomenological
theory. Under the proximity effect, the electric current in the normal metal
may be described by two contributions:
$\boldsymbol{j}= \boldsymbol{j}_N+\boldsymbol{j}_S$
with the normal current $\boldsymbol{j}_N=\sigma_0 \boldsymbol{E}$ and
the supercurrent $\boldsymbol{j}_S= - \frac{n_s e^2}{m c}\boldsymbol{A}
=i \frac{n_s e^2}{m\omega} \boldsymbol{E}$ with 
 $m$ being the mass of electron~\cite{trunin}.
The complex conductivity then becomes
$\sigma_1+i\sigma_2= \sigma_0+i \frac{n_s e^2}{m\omega}$.
The positive value of $n_s$ is indispensable for
explaining the Meissner effect. This argument can be easily confirmed when we
calculate the static magnetic field by solving the current $\boldsymbol{j}_S$ 
with $\nabla\times \boldsymbol{H}= \frac{4\pi}{c} \boldsymbol{j}_S$.
In the case of odd-frequency pairs, however,
$n_s$ becomes negative, which implies absence of the Meissener effect.
This fact also significantly modifies the impedance of a NS bilayer 
defined by
\begin{align}
Z_{\textrm{NS}}=&R_{\textrm{NS}}\!-\!iX_{\textrm{NS}}=\bar{Z}_{\textrm{N}} \frac{Z_{\textrm{S}} \cos \bar{k}_n L\! 
-\! i \bar{Z}_{\textrm{N}} \sin \bar{k}_n L }
{ \bar{Z}_{\textrm{N}} \cos \bar{k}_nL\! -\! iZ_{\textrm{S}} \sin \bar{k}_nL },\label{zns1}
\end{align}
where $Z_S$ is the impedance of the superconductor shown in Fig.~\ref{fig1}(a).
Since $L$ being comparable to $\xi_{T_c}$, the conductivity depends on $x$ only slightly
in the normal metal, which enables us to define spatially averaged values: 
the conductivity $\bar{\sigma}_{\textrm{N}} =  \int_{-L}^{0} dx \sigma_{\textrm{N}}(x)/L$, 
the impedance $\bar{Z}_{\textrm{N}}= -i \sqrt{ 4\pi i \omega / (c^2 \bar{\sigma}_{\textrm{N}})}
=\bar{R}_{\textrm{N}}-i\bar{X}_{\textrm{N}}$, 
and the wavenumber of 
electromagnetic field $\bar{k}_n=  \sqrt{ i {4\pi \omega \bar{\sigma}_{\textrm{N}}}/{c^2}  }$. 
The sign change of $\sigma_2$ results in the real value of $\bar{k}_n$ 
reflecting the absence of the Meissener effect.
In Fig.~\ref{fig3}, we show the impedance of a NS bilayer as a function of 
temperature for $s$-wave in (e) and chiral $p$-wave NS junctions in (f), 
where we assume the resistivity of a normal metal $\rho=\sigma_0^{-1}$ is about 
$2 \times 10^{-6} [\Omega \cdot \textrm{cm}]$.
In the $s$-wave symmetry, the impedance of the NS bilayer (e) shows the
conventional relation $R_{\textrm{NS}} < X_{\textrm{NS}}$. In the chiral $p$-wave symmetry, 
on the other hand, $R_{\textrm{NS}}$ goes beyond $X_{\textrm{NS}}$ at low temperature. 
In addition, the absence of the Meissener effect in the normal metal causes 
the sign change of $X_{\textrm{NS}}$ in the present results. 
Using relations $\bar{X}_{\textrm{N}}\ll \bar{R}_{\textrm{N}}$ and 
$Z_{\textrm{S}} \approx -iX_{\textrm{S}}$, it is possible to show $X_{\textrm{NS}} \approx 0$
at $\bar{R}_{\textrm{N}}/X_{\textrm{S}}=\tan\{ \textrm{Re}(\bar{k}_nL)\}$.

Finally, we briefly discuss realistic structures for observing the unusual relation $R>X$.
In experiments, it is possible to measure the impedance of a bilayer. 
One example is a bilayer consisting of a normal metal and Sr$_2$RuO$_4$~\cite{maeno}.
Since the pairing symmetry in Sr$_2$RuO$_4$ is spin-triplet chiral-$p$,
the anomalous relation $R>X$ as shown in Fig.~\ref{fig3}(f) would be observed in experiments.
At the same time, since our qualitative result is based on the symmetry of the induced 
superconducting state, the proposed effect can also be observed in conventional 
superconductor/ferromagnet structures where the odd-frequency state is also 
generated and can be spatially separated \cite{bergeret}. 
An example of this type is a bilayer of Nb/CrO$_2$, where only odd-frequency 
pairs can penetrate
into the half-metallic ferromagnet CrO$_2$~\cite{Keizer,ya07sfs,braude,eschrig}.


In summary, we have theoretically studied the impedance ($Z=R-iX$)
of a bilayer consisting of a normal metal and a superconductor.
For spin-singlet superconducting proximity structures, we
found the conventional and robust relation $R \ll X$.
For spin-triplet ones, $R$ can be larger than $X$  
and $X$ may change sign as a function of temperature at low $T$.
On the basis of obtained results, we propose a method to analyze
the spin-symmetry of Cooper pairs in a superconductor and to detect
odd-frequency superconductivity.

This work was supported by KAKENHI on Innovative Areas 
``Topological Quantum Phenomena''(No. 22103002).
Ya.V.F. was supported by the RFBR (Grant No.\ 11-02-00077-a),
the Russian Federal Agency of Education and Science (Contract
No.\ P799), and the program ``Quantum physics of condensed matter'' of the RAS.

\end{document}